\documentstyle[twoside,12pt]{article}
\newcommand{\dfrac}{\displaystyle \frac}
\newcommand{\nb}[1]{\mbox{\normalsize #1}}

\def\ao{{}\kern-.10em\hbox{``}}
\begin{document}
\large
\bibliographystyle{plain}

\begin{titlepage}
\hfill \begin{tabular}{l} HEPHY-PUB 596/94\\ UWThPh-1994-02\\ January 1994
\end{tabular}\\[4cm]
\begin{center}
{\Large\bf COMMENT ON \ao ANALYTIC SOLUTION OF THE RELATIVISTIC COULOMB
PROBLEM FOR A SPINLESS SALPETER EQUATION"}\\
\vspace{1.5cm}
{\Large\bf Wolfgang LUCHA}\\[.5cm]
{\large Institut f\"ur Hochenergiephysik\\
\"Osterreichische Akademie der Wissenschaften\\
Nikolsdorfergasse 18, A-1050 Wien, Austria}\\[1cm]
{\Large\bf Franz F. SCH\"OBERL}\\[.5cm]
{\large Institut f\"ur Theoretische Physik\\
Universit\"at Wien\\
Boltzmanngasse 5, A-1090 Wien, Austria}\\[1.5cm]
{\bf Abstract}
\end{center}
\normalsize

We demonstrate that the analytic solution for the set of energy eigenvalues
of the semi-relativistic Coulomb problem reported by B. and L. Durand is in
clear conflict with an upper bound on the ground-state energy level derived
by some straightforward variational procedure.
\end{titlepage}

Some time ago the authors of Ref. \cite{durand83} made an attempt to find the
exact analytic solution to the quantum-theoretic bound-state problem defined
by the semi-relativistic (since it incorporates fully relativistic kinematics
by involving the \ao square-root" operator of the relativistic kinetic
energy, $\sqrt{\vec p\,{}^2 + m^2}$) Hamiltonian corresponding to the
so-called \ao spinless Salpeter equation," for two particles of equal mass
$m$ given by
\begin{equation}
H = 2\sqrt{\vec p\,{}^2 + m^2} + V(\vec x) \quad ,
\label{eq:semrelham}
\end{equation}
for the special case where the arbitrary coordinate-dependent static
interaction potential $V(\vec x)$ is specified to be identical to the Coulomb
potential
\begin{equation}
V(r) = - \dfrac{\kappa}{r} \quad ,
\label{eq:coulpot}
\end{equation}
depending only on the radial coordinate $r \equiv |\vec x|$ and parametrized
by some coupling strength $\kappa$. In particular, for vanishing orbital
angular momentum $\ell$, i.~e., for $\ell = 0$, an analytic expression for
the totality of energy eigenvalues $E_n$ corresponding to the
semi-relativistic Coulomb problem has been derived, which reads
\cite{durand83}
\begin{equation}
E_n = \dfrac{2\,m}{\sqrt{1 + \dfrac{\kappa^2}{4\,n^2}}} \quad , \qquad
n = 1,2,\dots \quad .
\label{eq:e-durand}
\end{equation}

In this Comment, however, we take the liberty to express our rather severe
doubts on the general validity of this expression. Although we are, at
present, not able to give the correct energy eigenvalues, we are convinced
that Eq. (\ref{eq:e-durand}) must necessarily be wrong.

Our line of reasoning makes use of a standard variational technique in order
to derive a strict upper bound on the energy eigenvalue of the ground state.
For the case of the Coulomb potential (\ref{eq:coulpot}), this upper bound
turns out to be violated by the result obtained from Eq. (\ref{eq:e-durand})
for $n = 1$. The basic idea of this variational technique is
\begin{itemize}
\item to calculate the expectation values of the Hamiltonian $H$ under
consideration with respect to a suitably chosen set of trial states
$|\lambda\rangle$ distinguished from each other by some variational parameter
$\lambda$, which yields the $\lambda$-dependent expression $E(\lambda) \equiv
\langle\lambda|H|\lambda\rangle$, and
\item to minimize $E(\lambda)$ with respect to $\lambda$ in order to obtain
the upper bound to the proper energy eigenvalue $E$ of the Hamiltonian $H$ in
the Hilbert-space subsector of the employed trial states $|\lambda\rangle$ as
the above $\lambda$-dependent expression $E(\lambda)$ evaluated at the point
of the minimizing value $\lambda_{\nb{min}}$ of the variational parameter: $E
\le E(\lambda_{\nb{min}})$.
\end{itemize}
For the Coulomb potential, the most reasonable choice of trial states is
obviously the one for which the coordinate-space representation $\psi(\vec
x)$ of the states $|\lambda\rangle$ for vanishing radial and orbital angular
momentum quantum numbers is given by the hydrogen-like trial functions
($\lambda > 0$)
$$
\psi(\vec x) = \sqrt{\dfrac{\lambda^3}{\pi}}\,\exp(- \lambda\,r) \quad .
$$
For this particular set of trial functions we obtain for the expectation
values we shall be interested in, namely, the ones of the square of the
momentum $\vec p$ and of the inverse of the radial coordinate $r$,
respectively, with respect to the trial states $|\lambda\rangle$
$$
\left\langle\lambda\left|\vec p\,{}^2\right|\lambda\right\rangle = \lambda^2
$$
and
$$
\left\langle\lambda\left|\dfrac{1}{r}\right|\lambda\right\rangle = \lambda
\quad .
$$

Let us follow this line of arguments in some detail. As an immediate
consequence of the fundamental postulates of any quantum theory, the
expectation value of a given Hamiltonian $H$ taken with respect to any
normalized Hilbert-space state and therefore, in particular, taken with
respect to any of the above trial states must necessarily be larger than or
equal to that eigenvalue $E$ of the Hamiltonian $H$ which corresponds to its
ground state:
$$
E \le E(\lambda) \equiv \langle\lambda|H|\lambda\rangle \quad .
$$
Application to the semi-relativistic Hamiltonian of Eq. (\ref{eq:semrelham})
yields for the right-hand side of this inequality
$$
E(\lambda) =
2\left\langle\lambda\left|\sqrt{\vec p\,{}^2 + m^2}\right|\lambda\right\rangle
+ \langle\lambda|V(\vec x)|\lambda\rangle \quad .
$$

Here, the rather cumbersome although (for convenient trial states) not
impossible evaluation of the expectation value of the square-root operator
may be very easily circumvented by taking advantage of some trivial but
nevertheless fundamental inequality. This inequality relates the expectation
values, taken with respect to arbitrary Hilbert-space vectors $|\rangle$
normalized to unity, of both the first and second powers of a self-adjoint
but otherwise arbitrary operator ${\cal O} = {\cal O}^\dagger$; it reads
$$
|\langle{\cal O}\rangle| \le \sqrt{\langle{\cal O}^2\rangle} \quad .
$$
For the purposes of the present discussion it is sufficient to replace, in
turn, $E(\lambda)$ by its upper bound obtained by applying this inequality:
$$
E(\lambda) \le
2\sqrt{\left\langle\lambda\left|\vec p\,{}^2\right|\lambda\right\rangle + m^2}
+ \langle\lambda|V(\vec x)|\lambda\rangle \quad .
$$
Identifying in this---as far as its evaluation is concerned,
simplified---upper bound the up to now general potential $V(\vec x)$ with the
Coulomb potential (\ref{eq:coulpot}) and inserting both of the
$\lambda$-dependent expectation values given above implies
\begin{equation}
E(\lambda) \le 2\sqrt{\lambda^2 + m^2} - \kappa\,\lambda \quad .
\label{eq:simpupp}
\end{equation}

{}From this intermediate result, by inspection of the limit $\lambda\to\infty$,
we may state already at this very early stage that, for the semi-relativistic
Hamiltonian (\ref{eq:semrelham}), (\ref{eq:coulpot}) to be bounded from below
at all, the Coulombic coupling strength $\kappa$ has to stay below a certain
critical value: $\kappa \le 2$.

The value of the variational parameter $\lambda$ which minimizes the upper
bound on the right-hand side of Eq. (\ref{eq:simpupp}) may be determined from
the derivative of this expression with respect to $\lambda$:
$$
\lambda_{\nb{min}} = \dfrac{m\,\kappa}{2\sqrt{1 - \dfrac{\kappa^2}{4}}}
\quad .
$$
For this value of $\lambda$, by shuffling together all our previous
inequalities, we find that the energy eigenvalue corresponding to the ground
state of the semi-relativistic Hamiltonian (\ref{eq:semrelham}) with Coulomb
potential (\ref{eq:coulpot}), $E$, is bounded from above by
\begin{equation}
E \le 2\,m\sqrt{1 - \dfrac{\kappa^2}{4}} \quad .
\label{eq:groundupp}
\end{equation}
However, confronting this finding with the energy eigenvalue obtained from
Eq. (\ref{eq:e-durand}) for the lowest conceivable value of the quantum
number $n$, that is, $n = 1$,
$$
E_1 = \dfrac{2\,m}{\sqrt{1 + \dfrac{\kappa^2}{4}}} \quad ,
$$
we observe that, for any nonvanishing value of the coupling constant
$\kappa$, the inequality (\ref{eq:groundupp}) cannot be satisfied by the
ground state of Ref. \cite{durand83}. Consequently, we arrive at the already
announced conclusion that, unfortunately, there must be something wrong with
the analysis of the semi-relativistic Coulomb problem reported in Ref.
\cite{durand83}, in particular, with the quoted analytic solution for the
corresponding set of energy eigenvalues.

\end{document}